%Paper: hep-ph/9401303
%From: mende@het.brown.edu (Paul F. Mende)
%Date: Fri, 21 Jan 94 03:25:34 EST
%Date (revised): Sun, 23 Jan 94 00:26:13 EST

\input harvmac
\def\INSERTFIG#1{#1}

%% Figures:  There are 6 postscript figures packed with "uufiles,"
%% including unpacking instructions.  See the end or "part 2."
%%
%% If you don't want them embedded,
%% uncomment the next line and delete the one after that.

%\def\INSERTFIG#1{}   % uncomment to stop figure insertion
\input epsf

\def\CAPTION#1#2{{\narrower\noindent
\multiply\baselineskip by 3
\divide\baselineskip by 4
{\ninerm #1}{\ninesl #2 \medskip}}}

% Miscellaneous macros

\def\A{{\cal A}}
\def\B{{\cal B}}
\def\Nc{{N_c}}
\def\gpiBB{g_{\pi B n}}
\def\gpiBBo{g_{\pi B B}}
\def\gn{{f_{n}}}

\def\tL{\tilde\Lambda}
\def\d{{\rm d}}
\def\qm{q^2_{\rm max}}
\def\o{\omega}
\def\voo{\left({v\over\o}\right)}

\def\vmo{\left({v-\o\over1-\o}\right)}

\def\PI{\pi}
\def\FIGURE#1{}

\def\art#1{{\sl ``#1''}}

\def\pvint{-\mskip-19mu\int}
\def\d{{\rm d}}

\def\dy{{\rm d}y\,}

\def\PHI#1#2{\phi_{#1}( {#2} )}
\def\GAMMA#1#2{\Phi_{#1}( {#2} )}
\def\OMIT#1{}

\def\vev#1{{\left\langle #1 \right\rangle}}
\def\D{{\rm d}}

\def\ie{{\it i.e.}}
\def\eg{{\it e.g.}}

\def\CTP{{\it Comm.\ Theor.\ Phys.\ }}

\def\NC{{\it Nuovo Cimento\ }}
\def\NP{{\it Nucl.\ Phys.\ }}
\def\PL{{\it Phys.\ Lett.\ }}
\def\PR{{\it Phys.\ Rev.\ }}

\def\PRL{{\it Phys.\ Rev.\ Lett.\ }}
\def\SJNP{{\it Sov.~J.~Nucl.~Phys.\ }}

\def\ctp#1#2#3{\CTP{\bf #1} (#2) #3}
\def\nc#1#2#3{\NC{\bf #1} (#2) #3}
\def\np#1#2#3{\NP{\bf #1} (#2) #3}
\def\pl#1#2#3{\PL{\bf #1} (#2) #3}
\def\prl#1#2#3{\PRL{\bf #1} (#2) #3}
\def\pr#1#2#3{\PR{\bf #1} (#2) #3}
\def\sjnp#1#2#3{\SJNP{\bf #1} (#2) #3}

\def\authornote{\footnote}

%% References

\lref\faustov{R.N. Faustov, V.O. Galkin and A.Yu. Mishurov,
   \art{Semileptonic Decays Of D Mesons And B Mesons In The Relativistic Quark
Model,} \sjnp{55}{1992}{608}}%rel-QM

\lref\jaus{W. Jaus,
\art{Semileptonic Decays Of B And D Mesons
In The Light Front Formalism,}
\pr{D41}{1990}{3394}}%rel-QM

\lref\hagiwara{K. Hagiwara, A. D. Martin and M. F. Wade,
\art{Exclusive Semileptonic B Meson Decays,}
 \np{B327}{1989}{569}}%unavailable

\lref\bareiss{A. Bareiss and E. A. Paschos,
\art{Semileptonic B Meson Decays In The Parton Model,}
\np{B327}{1989}{353}}%unavailable

\lref\kizukuri{Y. Kizukuri,
\art{Semileptonic Decays Of B Meson And Effects Of Form-Factor,}
 \pl{109B}{1982}{93}}%unavailable

\lref\casalbuoni{R. Casalbuoni, A. Deandrea, N. Di Bartolomeo, R. Gatto, F.
Feruglio and G. Nardulli,
\art{Effective Lagrangian for Heavy and Light Mesons: Semileptonic Decays, }
\pl{B299}{1993}{139},
hep-ph/9211248}%PD+1pt

\lref\colangelo{P. Colangelo and P. Santorelli,
  \art{Dependence of the Form Factors of $B\to \pi\ell\nu$
  on the Heavy Quark Mass,}
  hep-ph/9312258, BARI-TH/93-163, DSF-T-93/45, INFN-NA-IV-93/45}

\lref\latticea{C. W. Bernard, A. X. El-Khadra and A. Soni,
\art{Lattice Approach to Semileptonic Decays of Charm Mesons,}
\np{B26 (Proc.Suppl.)}{1992}{204}}%PD-com1pt

\lref\latticeaa{C. W. Bernard, A. X. El-Khadra and A. Soni,
\art{Lattice Calculation Of The Semileptonic Form-Factor At The End Point,}
\pr{D47}{1993}{998}}%PD-com1pt

\lref\latticeb{V. Lubicz, G. Martinelli, M. S. McCarthy and C. T. Sachrajda,
\art{Semileptonic Decays Of D Mesons In A Lattice QCD,}
\pl{B274}{1992}{415}}%PD-com1pt(guess

\lref\latticebb{A. Abada et al,
\art{Semileptonic Decays Of Heavy Flavors On A Fine Grained Lattice,}
LPTENS-93-14, hep-lat/930800}%PD-com1pt(guess

\lref\guoa{X-H. Guo and T. Huang,
\art{Hadronic Wave Functions In D And B Decays,}
\pr{D43}{1991}{2931}}%must check

\lref\guob{X-H. Guo and T. Huang,
\art{Pseudoscalar Meson Wave-Functions And Exclusive Decays Of D And B
Mesons, }
\ctp{13}{1990}{365}}%unavailable

\lref\nussinov{S. Nussinov and W. Wetzel,
\art{Comparison Of Exclusive Decay Rates For $b\to u$ And $b\to c$
Transitions,}
 \pr{D36}{1987}{130}}%QM

\lref\gsw{B. Grinstein, M. B. Wise and N. Isgur,
\art{Weak Mixing Angles From Semileptonic Decays Using The Quark Model,}
\prl{56}{1986}{298}}%QM

\lref\isgw{N. Isgur, D. Scora, B. Grinstein and M. B. Wise,
\art{Semileptonic B And D Decays In The Quark Model,}
\pr{D39}{1989}{799}}%QM

\lref\ward{B.F.L. Ward,
\art{Lepage-Brodsky Approach To Semileptonic D Decay,}
\nc{98A}{1987}{401}}%QM

\lref\bauer{M. Bauer and M. Wirbel,
\art{Form-Factor Effects In Exclusive D And B Decays,}
Z. Phys. C42 (1989) 671 }%PD

\lref\wsb{M. Wirbel, B. Stech and M. Bauer,
\art{Exclusive Semileptonic Decays Of Heavy Mesons,}
Z. Phys. C29 (1985) 637}%PD

\lref\ball{P. Ball, V. M. Braun and H. G. Dosch,
  \art{Form-Factors Of Semileptonic D Decays From QCD Sum Rules,}
\pr{D44}{1991}{3567}}%QCD-SR

\lref\balltoo{P.Ball,
   \art{The Semileptonic Decays $D\to \pi(\rho) e \nu$ AND $B\to \pi
   (\rho) e \nu$ From QCD Sum Rules,}
hep-ph/9305267, TUM-T31-39/93}

\lref\CCG{C.~G.~Callan, N.~Coote, and D.~J.~Gross,
  \art{Two-dimensional Yang-Mills theory: a model of quark confinement,}
        \PR {\bf D13} (1976) 1649}

\lref\einhorn{M.~B.~Einhorn,
        \art{Confinement, form factors, and deep-inelastic scattering in
        two-dimensional quantum chromodynamics,}
        \PR {\bf D14} (1976) 3451}

\lref\thooft{G.~'t~Hooft,
        \art{A two-dimensional model for mesons,}
        \NP {\bf B75} (1974) 461}

\lref\JM{R.~L.~Jaffe and P.~F.~Mende,
	\art{When is field theory effective?}
	\NP {\bf B369} (1992) 189}

\lref\IW{N.~Isgur and M.~B.~Wise,
	\art{Influence of the $B^*$ resonance on
	$\bar B\to \pi e \bar\nu_e$,}
	\PR {\bf D41} (1990) 151}

\lref\wise{M.~B.~Wise,
	\art{Chiral perturbation theory for hadrons containing
	a heavy quark,}
	\PR {\bf D45} (1992) 2188}

\lref\falkluke{A.F. Falk and M. Luke,
   \art{Strong Decays of Excited Heavy Mesons
   in Chiral Perturbation Theory,}
   \pl{B292}{1992}{119}, hep-ph/9206241}

\lref\BDtoo{G.~Burdman and J.~F.~Donoghue,
   \pl{B280}{1992}{287}}

\lref\yanetal{T.~M.~Yan {\it et. al.\/}
   \pr{D46}{1992}{1148}}

\lref\BD{G.~Burdman and J.~F.~Donoghue,
	\art{Two component semileptonic form factors,}
	\PRL {\bf 68} (1992) 2887}

\lref\burd{G.~Burdman, Z. Ligeti, M. Neubert, and Y. Nir,
   \art{The Decay $B\to\pi\ell\nu$ in Heavy Quark Effective Theory,}
   hep-ph/9309272}

\lref\EXACT{B.~Grinstein and P.~F.~Mende,
	\art{Exact heavy to light meson form factors in the combined
	heavy quark, large $\Nc$ and chiral limits,}
	hep-ph/9312353, Brown-HET 928}

\lref\georgi{H.M.~Georgi,
   \art{On-shell effective field theory,}
    \NP {\bf B361} (1991) 369 }

\def\DATE{January 1994}

\Title{
\vbox{
\hbox{hep-ph/9401303}
\hbox{SSCL--Preprint--549}
\hbox{Brown HET--930}
}}
{\vbox{
  \centerline{Form Factors in the Heavy Quark and Chiral Limit:}
  \centerline{Pole Dominance in $\bar B\to \pi e \bar\nu_e$}
}}

\centerline{Benjam\'\i n Grinstein\authornote{$^\star$}%
{Address until February 9, 1994.
Research supported in part by the Alfred P. Sloan Foundation and by the
U.S. Department of Energy under contract DE--AC35--89ER40486.
E-mail: {\tt ben@smuphy.physics.smu.edu}}
and Paul F.~Mende\authornote{$^{\star\star}$}%
{Research supported by an SSC Fellowship, TNRLC \#FCFY9322,
and by the U.S. Department of Energy under grant DE-AC02-76-ER03130.
E-mail: {\tt mende@het.brown.edu}}}
\medskip
\centerline{\sl $^\star$Superconducting Super Collider Laboratory,
Dallas, Texas 75237}
\smallskip
\centerline{\sl $^{\star\star}$Department of Physics, Brown University,
Providence, Rhode Island 02912}

\bigskip\bigskip

It is shown that the form factors in the semileptonic decay of
ground state heavy
mesons to light pseudoscalar mesons are dominated by the vector meson
pole at all momentum transfers.
First, the general approaches to modeling form
factors in terms of quark model, pole dominance, or hybrid shapes are
reviewed.  It is then shown that in the combined limits of heavy
quarks, chiral symmetry, and large~$\Nc$ the form factor is precisely
pole-dominated. It is also shown that in a two-dimensional QCD model
pole-dominance occurs for any heavy quark mass.
Corrections to this limit are discussed and, to illustrate the nature
of the approximations, an explicit exact calculation of the approach
to the limit in a two-dimensional model is given.

\Date{\DATE}

\eject

\newsec{Introduction}

Weak decays of heavy hadrons directly probe
flavor changing interactions.
In the standard model of electroweak interactions these decays
provide a means of extracting fundamental parameters of the theory:
four mixing angles, $|V_{c d}|$, $|V_{c s}|$, $|V_{u b}|$ and $|V_{c b}|$,
which determine the decay rates of the heavy hadrons.
In practice, however,
our inability to compute these rates from first principles limits
efforts to extract these fundamental angles.
Even when limiting our attention to semileptonic
decays, the simpler hadronic matrix elements prove too
difficult to compute.

In processes involving the semileptonic decay of a heavy quark
to another heavy quark, as in $b\to c\ell\nu$, a systematic
expansion in powers of the hadronic scale over the heavy masses
allows computation of the hadronic matrix elements involved at one
kinematic end-point, namely, the point at which the resulting
hadronic system is at rest in the rest frame of the original hadron.
This ``heavy quark effective theory'' (HQET) is therefore of great
relevance to the determination of $|V_{c b}|$.

But for decays of a heavy quark into a {\it light\/} one,
there is unfortunately little that the HQET has to say.
While one can find relations between different
measurable decays, it does not seem possible to calculate  the
matrix elements themselves.

In view of the lack of first-principles calculations of
heavy-to-light decay rates,  phenomenological estimates depend on
hadronic models used. It is no surprise that the results vary
considerably.

Present hadronic model
calculations of heavy-to-light decay rates attempt first to estimate
the hadronic matrix element at one kinematic point and then
extrapolate based on some assumed functional shape of the form factor,
chosen in a somewhat ad-hoc manner.
This `shape' is one of three types:

\medskip
\item{({\it i})}{\it Pole-Dominated\/} shapes assume the form, although not
necessarily
the residue,  of a form factor dominated by the lowest-lying
resonance that couples to the weak current, \eg,
vector-meson dominance of the $B^*$ in $b\to u$
transitions\refs{\wsb\bauer\latticebb\latticeb\latticeaa\latticea%
\casalbuoni{--}\colangelo}.

\item{({\it ii})}{\it Quark-Model\/} shapes are suggested by an extrapolation
of the non-relativistic constituent quark model from the low recoil
region to the relativistic region.  The extrapolation is not unique
since the non-relativistic computation does not automatically yield a
Lorentz-invariant form factor\refs{\nussinov\gsw{--}\isgw}.
Relativistic wavefunctions have also been used to
compute the shape\refs{\faustov\jaus\guoa\guob{--}\ward}.
\item{({\it iii})}{\it Two-Component\/} models assume a linear combination
of the two preceding shapes\refs{\IW\BD\ball{--}\balltoo}.
\medskip

{}From the practical point of view this situation is quite inadequate;
from a theoretical one it is far worse:
it leaves us with no insight into the dominant structures and dynamics
of these hadrons.
More optimistically, detailed knowledge of the correct~$q^2$
dependence promises both to help in the experimental
determination of fundamental parameters
and to bridge the variety of theoretical ideas
underlying the present matrix element calculations.

In this paper we  argue that there is compelling evidence in favor
of the first, {\bf pole-dominated,} structure for this form factor.
We show that the $q^2$ dependence is of the single-pole form
in the combined heavy quark, large~$\Nc$ and chiral limit, where
the hadronic matrix element is exactly computed, as we have
discussed in a recent letter\EXACT.
We go on to study decays of heavy mesons in two-dimensional planar QCD,
where we find striking confirmation of both the general results
and the expected approach to the limiting case.
While the 't~Hooft model is in no sense a controlled approximation to
the four-dimensional world, it offers valuable insights.
Most importantly it provides a testing grounds for new models
since, as we explain, many considerations motivating the
modeling of form factors are independent of those details (dimension
and number of colors) which make the model solvable.

We have also discovered, in the course of this work, a number of useful
relations that permit calculation of some matrix elements in this model
for the first time.
Thus we expect these results to be interesting as well
to those interested in two-dimensional field theory for its own sake.

The paper is organized as follows.
In section~2 we discuss  physical arguments
that suggest the validity of the form factor shapes described above.
We pay particular attention to arguments
due to Isgur and Wise
that use simple power counting to argue in favor of a
two-component model\refs{\IW, \BD}.
In section~3 we prove pole dominance at large~$\Nc$.
This is done first
for the simpler two-dimensional case for which spin is absent and
the heavy quark limit need not be taken.
The four-dimensional generalization is then presented.
In section~4 we fleetingly review the 't~Hooft model in order
to introduce notations and conventions.
In section~5  we calculate weak decay form factors
and discuss the results in section~6.
The principal field-theoretic results
of the paper are contained in section~3 and section~5,
and the impatient reader may wish to jump there directly.

\newsec{Models of Form Factors}

Several possible shapes for
the form factors of heavy-to-light decays have been proposed
in the literature.
Always the idea is to {\it extrapolate\/} (or guess) a solid calculation of
a hadronic matrix elements at a {\it single\/}  kinematic point
to a form factor at other momentum transfers.

{\it Pole-Dominated\/} form factors are assumed to have the
functional form
   \eqn\polemodelffs{
   f_\pm(q^2)={C_\pm \over 1-q^2/{\mu^\star}^2} .
   }
where $\mu^\star$ is the mass of the lightest state which
couples to the weak current $V_\mu$.
The strength of the form factor~$C_\pm$  is  obtained from by
estimating the matrix element at one kinematic point $q^2=Q^2$:
   \eqn\Polestrength{
   C_\pm \equiv f_\pm(Q^2)\, \left( 1 - Q^2/{\mu^\star}^2 \right).
   }
In $B\to\pi$ transitions this pole belongs to the $B^*$ vector meson.
(In two dimensions where spin is absent the corresponding
state is the $B$-meson itself which couples directly to the vector
current).

Pole-dominated form factors also arise in the
chiral Lagrangian approach to heavy
quark interactions\refs{\wise\BDtoo\yanetal\falkluke{--}\burd};
however this analysis is necessarily limited to small momentum
transfers and applies not to a large range of $q^2$ but
only to the small-recoil regime in the vicinity of the $B^*$ pole.

{\it Quark-Model\/} forms
are suggested by extrapolating a non-relativistic functional shape
to the relativistic region.
This shape is computed in the low-recoil region
using the constituent quark model.
The  extrapolation is not unique since the non-relativistic
computation  does not automatically yield a Lorentz-invariant form factor.
The computation requires an overlap integral
of the non-relativistic
wave-functions for the ground state mesons.
For example, in the rest frame of the $B$-meson
\eqn\isgwmodelffs{
  \eqalign{
   (\mu_B+\mu_\PI) \tilde f_+ + (\mu_B-\mu_\PI) \tilde f_-
   &= \sqrt{4\mu_B\mu_\PI}\int\D \vec k \,
   \phi^*_\PI(\widetilde{\vec p} +\vec k) \phi_B(\vec k)\cr
   (\tilde f_+ - \tilde f_-)\vec p
   = \sqrt{4\mu_B\mu_\PI}&\int \D \vec k \,
   \phi^*_\PI(\widetilde{\vec p} + \vec k)
     \left( {\vec k\over 2m_b}+ {\vec k + \vec p \over 2 m_q}\right)
	\phi_B(\vec k) ,\cr
  }
}
where $\vec p$ is the spatial momentum of the $\PI$-meson,
$\widetilde{\vec p}\equiv (m_q/\mu_\PI)\vec p$,
$\phi_X$ is the non-relativistic wave-function of the $X$-meson,
and the tilde marks form factors which are not Lorentz-invariant.
By rotational  invariance, $\tilde f_\pm$ are functions of
$|\vec p|^2$ only.
One  may  construct Lorentz-invariant form factors
by  writing the momentum dependence in terms of a
reasonably chosen--- albeit
ad-hoc ---replacement
$|\vec p|^2 \to g(q^2)$, such  that
$g(q^2)\to |\vec p|^2$ as $q^2 \to \qm \equiv (\mu_B- \mu_\PI)^2$.
For example, since in the $B$-rest-frame
   \eqn\onesilly{
   q^2=\mu_B^2+\mu_\PI^2-2\mu_BE_\PI~,
   }
where $E_\PI=\sqrt{|\vec p|^2 + \mu_\PI^2}$, one may take
   \eqn\twosilly{
   g(q^2)
   \equiv \left({q^2-\mu_B^2-\mu_\PI^2\over2\mu_B}\right)^2-\mu_\PI^2
   }
or, alternatively
   \eqn\threesilly{
   g(q^2)\equiv [q^2-(\mu_B+\mu_\PI)^2]{\mu_\PI\over \mu_B}
   }

It is natural to suspect that these two approaches might be
quite limited and that at best each might be
applicable to a particular kinematic regime.

Indeed, a dispersion relation may be used to write the form factor as
a sum over contributions of resonances
   \eqn\ffsdisp{
   f_\pm(q^2) = {f_{B^*}g_{\pi B B^*}\over q^2-\mu^2_{B^*}} +
   \hat f_\pm(q^2),
   }
where $\hat f_\pm$ is the continuum contribution,\footnote{$^\star$}
{
The precise nature of the continuum contribution $\hat f$ is
not important, as
we need know only its structure in the complex $q^2$
plane,
and the strength of the associated singularity, but not the precise
nature of the singularity.
}
or single poles in the narrow
width approximation or large $\Nc$ limit:
   \eqn\contassum{
   \hat f(q^2) = \sum_{n>0} {f_{B^*} \gpiBB \over q^2-\mu^2_{n}}
   .}

The pole of course dominates near
the kinematic endpoint
when $|q^2-\mu_{B^*}^2|$ is much
smaller than the spacing between the $B^*$ and the next resonance
that couples to the current $V_\mu$.

By the same token, the lowest term
is not expected to dominate in the opposite
situation, when  $|q^2-\mu_{B^*}^2|$ is greater than the typical
spacing between resonances.
In this case many
resonances will in general contribute substantially to the form factor.
This is a signal that one is not expanding in a useful set of degrees of
freedom, and
a quark model description may be more adequate in this case.

{\it Two-component models\/} attempt to capture both these behaviors
particularly when one is interested in a large kinematic range.
The character of the actual transition is naturally of great interest.

While suggestive,
these simple arguments prove nothing.
Pole-dominance fails, for example, if the lowest
residue is anomalously small, so that the ``nearest singularity''
dominates only for
values of $|q^2-\mu_X^{\ast2}|$ so tiny that they are not in the
physical region.
It could happen that the higher states, often neglected,
actually have large residues which oscillate rapidly.
Form factors of this type in fact occur in QCD;
they are crucial to understanding transition between effective field
theories and quark model physics for $q\bar q$ states of
a single flavor\JM.
On the other hand the residue of the lowest lying
pole could be so much larger than all of the rest that dominance by
that pole alone could be guaranteed for all values of
$q^2 < \mu_X^{\ast2}$.

Isgur and Wise argued for
two-component	model in heavy-to-light decays\IW.
This has also been discussed in detail by Burdman and Donoghue\BD.
They argue that in the
combined chiral and heavy-quark limits
the form  factor is pole-dominated around $\qm$ --- and only there.
Now it is obvious that there is a $B^*$ pole and that this pole
plays a special role since the physical value of
$|\qm - \mu_{B^*}^2|$ is so small.

Consider the combined limit
\eqn\comblimit{
   \mu_\pi\to0 , \qquad M_b\to\infty ,
   \qquad \tL^2\equiv \mu_\pi \mu_B {\rm \ \ fixed}
}
(the analysis of Ref.~\IW,\BD, corresponds to
$\tL=0$).
Standard power counting gives
  $f_{B^*}\propto M_b^{1/2}$ and
  $g_{\pi B B^*} \propto M_b$.
On the other hand,
  $\mu_{B^*}^2- \mu_B^2\sim (M_b)^0$,
while for higher resonances ($n\ge1$) one has
  $\mu_n^2- \mu_B^2\propto M_b$.
At $\qm=(\mu_B-\mu_\pi)^2$
the denominator in \ffsdisp\ scales as~$M_b^0$,
while the denominators of $\tilde f_{\pm}$ (cf.~\contassum)
scale as~$M_b$.
In  the combined limit, the form factors at $\qm$ are pole dominated.
In fact the single pole dominates over a region that
scales like $\qm-q^2\propto M_b$.
This is a small fraction of the physical region,
which scales like~$M_b^2$.
Thus, Isgur and Wise conclude that a two-component form factor is
appropriate: the pole term dominates around~$\qm$
with residue given by
$f_{B^*}g_{\pi B B^*}$,
and the quark-model term is dominant at smaller $q^2$.

Though suggestive, this scenario rests on crucial assumptions.  The
key question is over what range the single pole dominates, and this
depends on dynamical calculations.  We will shortly examine these
assumptions critically in the case of a fully calculable toy model
which satisfies the same assumptions, yet displays form factors that
are truly pole-dominated over a much larger kinematic range.  It is
tempting to conjecture that this holds in four dimensions.  Before
describing our exact two-dimensional results, we explain a limit in
which the behavior holds exactly in four as well as two dimensions.

\newsec{Pole dominance at large $\Nc$}
The matrix element can be expressed in terms of two form factors~$f_\pm$:
\eqn\ffsdefd{
\vev{\PI(p') | V_\nu | \bar B(p)} = (p+p')_\nu f_+(q^2) + (p-p')_\nu f_-
(q^2)~.
}
where $V_\nu=\bar q\gamma_\nu Q$, $q=p-p'$
and throughout the paper $p$ and~$p'$
alway denote the momentum of the~$B$ and~$\pi$ mesons,
respectively.
The mesons $B$ and
$\PI$ have quantum numbers that correspond to the valence quarks
$Q\bar q$ and $q\bar q$, respectively.

Let us derive the pole-dominated shape of $\bar B\to\PI$ decay
in a suitable limit, and then discuss how the approximations might
be relaxed.

Consider QCD in the limit of large $\Nc$, with one heavy quark~$Q$ of
mass~$M$ and a light quark~$q$ of mass~$m$.
The form factors in \ffsdefd\ are saturated by couplings
of the flavor changing current~$V_\nu$ to the $Q\bar q$ resonances
in that channel.

To momentarily suppress the complication of spin, we pass to two
dimensions where we can write
   \eqn\Forms{
   \vev{\PI | V_\nu | B}
   = \sum_n {\vev{0|V_\nu|B_n}  \vev{\PI B_n|B}\over q^2 - \mu_n^2 }
   .}

For odd parity states $|B_n\rangle$,
   \eqn\oddfdef{
   \vev{0|V_\nu|B_n} = \epsilon_{\nu\lambda}q^\lambda \gn
   .}
We can describe these interactions conveniently in terms
an effective hadron Lagrangian,
\eqn\Ldef{
   {\cal L} = \sum_n \left( {1\over 2} (\partial_\lambda\varphi_n)^2
   - {\mu_n^2\over 2}\varphi_n^2 \right)
   + {\cal L}_{\rm int},
}
where a field $\varphi_n(x)$ is introduced for each meson state
and ${\cal L}_{\rm int}$ couples the mesons via terms
   \eqn\Lodd{
   {\cal L}_{\rm int}=
   \sum_{abc}
   \hat g_{abc}(q^2)\epsilon_{\lambda\nu}
   \, \partial_\lambda \varphi^a \partial_\nu \varphi^b \varphi^c~.
   }
Similarly,  for even parity,
\eqn\evenfdef{
   \vev{0|V_\nu|B_n} = q_\nu \gn
   ,}
with couplings
   \eqn\Leven{
   {\cal L}_{\rm int}=
   \sum_{abc}
    \hat g_{abc}(q^2) \, \varphi^a\varphi^b\varphi^c.
   }

The form factors of Eq.~\ffsdefd\ can then be written
\eqn\ffexpn{
   \eqalign{
   f_+(q^2) &= \sum_{{\rm even\ parity}}
   { \gn \hat\gpiBB(q^2) q^2 \over q^2-\mu_n^2}
   \cr
   f_-(q^2)&=\sum_{{\rm odd\ parity}} { \gn \hat\gpiBB(q^2) \over q^2-\mu_n^2}
   - \sum_{{\rm even\ parity}}
   { \gn \hat\gpiBB(q^2) (\mu_B^2-\mu_\pi^2) \over q^2-\mu_n^2}
   ~, \cr }
}
which is obtained with the help of the useful
two-dimensional formula
\eqn\useful{
   \epsilon_{\lambda\nu}q^\nu
   = \left\lbrack -q^2(p+p')_\lambda + (\mu_B^2 - \mu_\pi^2)(p-p')_\nu
   \right\rbrack / 2\epsilon^{\rho\sigma}p_\rho p'_\sigma ~.
}
Note that the expansions~\ffexpn\ have momentum dependent
numerators, proportional to the off-shell three point
couplings, $\hat\gpiBB(q^2)$.
Using analyticity and assuming suitable convergence, as will
be justified later, a contour integral gives
\eqn\ffexpntoo{
   \eqalign{
   f_+ (q^2)&= \sum_{{\rm even\ parity}}
   { \gn \hat\gpiBB(\mu_n^2) \mu_n^2 \over q^2-\mu_n^2}
   \cr
   f_- (q^2)&= \sum_{{\rm odd\ parity}}
   { \gn \hat\gpiBB(\mu_n^2) \over q^2-\mu_n^2}
   - \sum_{{\rm even\ parity}}
   { \gn \hat\gpiBB(\mu_n^2) (\mu_B^2-\mu_\pi^2) \over q^2-\mu_n^2}
   ~, \cr }
}
in which the numerators are the on-shell, physical couplings.

We make several observations.
It is large $\Nc$ which allows us to treat the resonances as stable
without continuum couplings in Eq.~\Forms.
It selects the three point couplings
to one-particle intermediate states.
In using convergence as $|q^2|\to \infty$,
we make an assumption about the large-momentum behavior
of the interactions, information unavailable from
a low-energy analysis or standard chiral Lagrangian analysis.
The shift of the numerators to the residues at the poles,
familiar in dispersion theory, has a simple physical origin:
in the ``effective'' meson field theory there is
freedom to make arbitrary field redefinitions without
changing the on-shell $S$-matrix.
Here that freedom is used to replace the momentum dependent couplings
(that is, the higher derivative operators)
by constants at the expense of
shifting the coefficients of higher point functions, which in
turn are down by powers of $1/N$.
We make such a shift.
Note the contrast with the use of field redefinitions by Georgi
in Ref.~\georgi, in which higher point functions were
suppressed instead by powers of the cutoff
to yield a related kind of ``on-shell effective theory.''

Now in the chiral limit --- the light quark mass $m\to 0$
and $\mu_\pi^2\to 0$ ---
the decay constants $f_{n}$ remain finite
while the on-shell three-point elements $\vev{\PI B_n|B}$ vanish.
This is a direct consequence of chiral symmetry and leads immediately
to a pole-dominated form factor.

To show this in detail, fix the state $B_n$ and consider
the matrix element of the {\it light-light\/} current
$a_\lambda=\bar q\gamma_\lambda \gamma_5 q$:
   \eqnn\Lightlight
$$ \eqalign{ \vev{B_n(q)| a_\lambda |B(p)} &= p'_\lambda
    F_n(p^{\prime2}) \cr
    &=\sum_\ell { \vev{ 0 | a_\lambda | \PI_\ell}
    \vev{\PI_\ell B_n|B} \over p'^2 - \mu_\ell^2 }\cr} $$
Here the sum is over the tower of $\bar qq$ states,
$|\PI_\ell\rangle$.  The form factors $F_n(p^{\prime2})$ can be written
in terms of the off-shell couplings
$\hat g_{\ell B n}(p'^2)$ and decay constants $f_\ell$
of the effective Lagrangian, Eqs.~\Lodd\ and~\Leven.
For parity odd $B_n$-states
\eqn\Foddgiven{
F_n(p^{\prime2})=\sum_\ell
  {f_\ell 2\epsilon^{\mu\nu}p_\mu p'_\nu\hat g_{\ell B n}(p'^2) \over
	 p'^2 - \mu_\ell^2 }
}
and for parity even states
\eqn\Fevengiven{
F_n(p^{\prime2})=\sum_\ell
  {f_\ell \hat g_{\ell B n}(p'^2) \over
	 p'^2 - \mu_\ell^2 }
}
Again, given suitable convergence, one may apply Cauchy's theorem to
these form factors and replace the numerators by the corresponding
on-shell expressions, \ie, $p'^2\to\mu^2_\ell$.

Next, axial current conservation implies
$f_\ell=0$ unless $\mu_\ell=0$,
so the axial current couples only to the massless
pion, $\PI=\PI_0$, and only the $\ell=0$ term persists in
Eqs.~\Foddgiven\ and~\Fevengiven\ in the chiral limit.

Again applying axial current conservation, $\partial\cdot a=0$, we have
   \eqn\Axial{
   \eqalign{
   0=p'^2 & F_n(p'^2)
 \cr
   &\to \cases{
   f_\pi \hat\gpiBB & if $n$ is parity even; \cr
   f_\pi \hat\gpiBB (2\epsilon^{\lambda\sigma}p_\lambda p'_\sigma)
    & if $n$ is parity odd,
   \cr}
   }}
for all three states on-shell. It immediately follows that
   \eqn\Vanish{
   \hat\gpiBB = 0,
   \qquad n \ne 0
   .}
so  all the coupling constants in the effective Lagrangian
\Lodd, \Leven\ except
to the ground state vanish on-shell.
The case $n=0$ is singled out because, on-shell,  the factor
$2\epsilon^{\lambda\sigma}p_\lambda p'_\sigma=0$, so it need
not follow that $g_{\pi B B}$ vanishes.  Indeed, it does not.

%%%

Combining these results and
introducing the coupling $\gpiBBo\equiv\mu^2_B\hat\gpiBBo(\mu_B^2)$ in
analogy with the definition that is natural in four dimensions,
the form factors are
\eqn\twodimffs{
   f_+(q^2)
   = - f_- (q^2)
   = { f_B \,\hat\gpiBBo(\mu_B^2) \mu_B^2\over q^2-\mu_B^2 }
   =   {f_B\,\gpiBBo\over q^2-\mu_B^2}
}

This readily generalizes to four dimensions, remaining
still in the limit of large~$\Nc$.
The detailed argument for the limiting case is given in Ref.~\EXACT.
The vector and axial-vector currents are of course no longer dual,
and the lightest state to couple to the $\pi B$ is the
$B^*$ vector meson, not the $B$ itself.
The essential point is that $B$ and $B^*$ are mass-degenerate
in the heavy quark limit so that the light-quark axial vector
current which produces pions can rotate the states into each other.

As above, look in the light-light channel
for a fixed state $B_n$
and take the divergence in a frame where
$\vec p\,' =0$:
   \eqn\FourD{
   0 = {p'}^\lambda \vev{B_n|a_\lambda|B}
   = \sum_\ell {p'}^0 {\vev{0|a_0|\PI_\ell}\vev{\PI_\ell B_n|B}
   \over p'^2 - \mu_\ell^2}
   = { \vev{\pi B_n|B} f_\pi{ p'}^2
   \over p'^2 - \mu_\pi^2}
   }
so that chiral symmetry again implies that
$\vev{\pi B_n| B}=0$
which in turn implies the vanishing of all $\gpiBB$
except for $n=0$.
The way in which this comes about is naturally different
in four dimensions, reflecting the corresponding spins,
symmetries and kinematics.
The states~$B_n$ must be analyzed according to their spin.
Only for spin one is there the possibility of a non-vanishing
coupling\EXACT.

One can write
   \eqn\Bstardecay{
   \vev{0| V_\lambda|B_n} = f_{n}\epsilon_\lambda
   ,}
where~$\epsilon_\lambda$ is the polarization of the~$B_n$
and then define couplings by
   \eqn\Bstarff{
   \vev{\pi B_n|B}
   = \epsilon^\lambda (p+p')_\lambda g^{(n)}_+
   + \epsilon^\lambda (p-p')_\lambda g^{(n)}_-
   .}

Nothing is learned about the~$g_-$ since $B_n$ is on-shell.
It is easy to see in the~$B_n$ rest frame
$\vec p = \vec p\,'$ that
\eqn\restgminus{
   g^{(n)}_- \vec\epsilon\cdot\vec p = 0
}
Let us restrict attention for now to the exact chiral limit
in the heavy quark (infinite) mass limit.
Then
$$
   \vec\epsilon\cdot\vec p  = |\vec p| \cos\theta ,
$$
where~$\theta$ is the angle between the polarization and the momentum
vectors, and is generally non-vanishing.
But from the kinematics it is also true that
\eqn\magofp{
 |\vec p| ={ \mu_n^2 - \mu_B^2 \over 2 \mu_n}
          =\cases{ \Lambda_0^2 / 2\mu_{B^*} = \CO(1/M_Q) & for $n = B^*$ ,\cr
                   \Lambda_n + {\cal O}(1/M_Q) & otherwise, \cr
                    }
}
where we introduce the mass difference $\Lambda_n\equiv \mu_n-\mu_B$ for
$n\neq B^*$ states, and take the large mass limit in the last equality.
Therefore, taking $M\to \infty$, the chiral limit then yields
\eqn\FINAL{
   0 = g_+^{(n)}\, |\vec p| =
         \cases{ 0 & for $n = B^*$ , \cr
                     g_+^{(n)}\Lambda_n & otherwise. \cr
                    }
}
So the couplings to excited states vanish, or
$g^{(n)}_+ / g^{(B^*)}_+ \to 0$.

What can we conclude from this?
We see that in four as in two dimensions, the effective
Lagrangian in the combined limit of $M\to\infty$, $m\to 0$,
and $\Nc\to\infty$ has only a single coupling so that the
form factors for semileptonic $B\to\pi$ decays are
{\bf necessarily pole-dominated in all kinematic regions}.
Moreover, in the large~$\Nc$ limit the behavior as the light
quark mass vanishes is expected to be smooth, so if we are not
strictly at the chiral limit, the corrections to the pole-dominated
form factor are ${\cal O}(\mu_\pi^2/\Lambda^2)$.

In order to make this point abundantly explicit, we
compute the form factors and effective Lagrangian coupling
to {\it all states\/} below in the 't~Hooft model
and explicitly compute all terms in Eqs.~\Forms, \Lodd, \Leven.
Not only will this illustrate the proof given here,
but it allows analysis of
the corrections to the limit when $m\ne 0$, and study of the
dependence on the heavy mass $M$.

Mesons may be classified as heavy or light in two dimensions much as in four.
In four dimensions the QCD Lagrangian is renormalizable and masses are
considered according to whether they are large or small compared to the
scale $\Lambda_{\rm QCD}$.
In two dimensions, QCD is super-renormalizable but the gauge coupling,
which has dimensions of mass, plays a role analogous to $\Lambda_{QCD}$
and serves to separate heavy from light.

It is equally clear that these arguments no longer apply when
the final state meson is no longer the near-massless ground state.
It is of great interest, therefore, to explore the decays
$B\to \pi'$, $B\to\rho$.

\newsec{The 't~Hooft model}

This model has been extensively studied and our work relies
on technology pioneered by 't~Hooft\thooft,
Callan, Coote, and Gross\CCG, and Einhorn\einhorn.
In these papers the bound state equations were derived; and
it was shown that the scattering amplitudes---and the form factor
in particular---can be written entirely in terms of interactions
among the meson bound states, with no quarks in the spectrum
or in the singularity structure of the amplitudes.

We recall the features of the model which
make it solvable,
and refer the reader to the
original papers for details.
The dynamics are defined by the Lagrangian,
  \eqn\YMLagrangian{
   {\cal L} = {1\over 4} \tr\, F^2
   + \sum_{a} 	\bar\psi_a(\gamma^\mu(i\partial_\mu - g_0 A_\mu)  -m_a)\psi_a
   ,
   }
where $A_\mu$ is an $SU(N)$ gauge field, $F_{\mu\nu}$ is its field strength
and $\psi_a$ is a Dirac fermion of mass~$m_a$.
In the large-$\Nc$ limit, the gauge coupling is scaled with $\Nc$:
\hbox{$g^2 = g_0^2N$} is held fixed as $\Nc\to\infty$.
The label $a$ runs over two flavors of quark, with
bare masses $m$ and $M$.

We relate this Yang-Mills Lagrangian~\YMLagrangian\
to the effective meson theory hadronic Lagrangian~\Ldef\
by the most pedestrian, concrete method imaginable:
by computing the $S$-matrix elements of physical states
in the model~\YMLagrangian\ and identifying directly with
the coupling functions in~\Ldef\ which reproduce the
same physics.

The main obstacle is that there are no known analytic solutions
to the bound state equations, and we have found neither approximation
techniques nor limiting cases that adequately serve us in
the most interesting regime.
We therefore turn to numerical techniques.
We compute the bound state wave functions and from them
evaluate the form factor as well the precise couplings
of the low-energy effective field theory when the quarks are
``integrated out.''
The techniques we use are those developed in Ref.~\JM.

The theory is most conveniently quantized in light-cone gauge.
Because there are no transverse dimensions,
setting~$A_-=0$ eliminates the gluon self-coupling.
It also serves to project gamma matrices onto a single component in
any Feynman graph that has just gluon vertices and ($-$) component
current insertions on fermion lines.
The infrared divergence in the gluon propagator, $i/k_-^2$, is regulated
by taking the principal value at the pole.

The leading term of the $1/\Nc$ expansion is the sum of planar graphs.
The equations for the full
propagator and self-energy can be solved exactly, with an
extremely simple result:
the net effect of all gluons starting and ending on the same
fermion line is just to renormalize the quark mass appearing in the
propagator,
   \eqn\noname{
   m^2 \to  \tilde m^2 \equiv m^2 - g^2/\pi
   ,}
so the full quark propagator is
   \eqn\Sdef{
   S(k) = {i k_- \over k^2 -  \tilde m^2  +i\epsilon}
   %= {i\over 2k_+ - {( M^2 -i\epsilon ) / k_-}}
   .}

After making this shift, all remaining gluon interactions enter as
ladder-type exchanges.
Crossings would require either gluon self-couplings,
which are absent,
or non-planar graphs, which are higher order in~$1/\Nc$.
The Yang-Mills coupling constant~$g$ has dimensions of mass, and
we choose units such that $g^2/\pi =1$, leaving
$m^2$ as the single dimensionless number parameterizing
the theory.
As is well known, there is a discrete spectrum of free mesons.

$\GAMMA{}{p,q}$ is the full meson-quark vertex, and the
wavefunction $\PHI{}{p,q}$ is defined by
   \eqn\noname{
   \PHI{}{p,q} = {1\over i\pi}\int\d p_+\, \GAMMA{}{p,q}S(p)S(p-q)
   .}
$\phi$ and $\Phi$
only depend on $p$ through the variable
$x = p_-/q_-$, so we denote
$\PHI{}{x} \equiv \PHI{}{p,q}$ and $\GAMMA{}{x} \equiv \GAMMA{}{p,q} q_-$.
In terms of $\phi$, the bound state equation
may be written as
   \eqn\BS{
   \mu_n^2 \PHI{n}{x}
   = \left({ M^2-1  \over x} + {m^2-1 \over 1-x}\right)\PHI{n}{x} -
   \pvint_0^1 {\dy\over (y-x)^2} \PHI{n}{y}
   .}
Here
$\PHI{n}{x}$ is a light-cone momentum space wavefunction of the
$n^{th}$ eigenstate, with mass~$\mu_n$,
and~$x=p_-/q_-$ is the fraction of light-cone momentum carried by the heavy
quark.
The $\PHI{n}{x}$ vanish at the boundaries, and consistency of $\BS$
requires that as $x\to 0$, $\PHI{n}{x} \sim x^\beta$, with
   \eqn\Betadef{
   \pi\beta_M\cot\pi\beta_M=1-M^2
   ,}
and similarly as $x\to 1$, as dictated
by the boundary behavior of the Hilbert transform.
This equation does not have solutions in terms of
known functions, but may be readily
solved numerically.

The range of $x$ for the bound states is always in the
interval~[0, 1], and $\phi=0$ outside of this range;
but~$\BS$ determines as well the
full meson-fermion-antifermion vertex,
   \eqn\vertexdef{
   \GAMMA{n}{z} = \int_0^1 {\dy\over (y-z)^2} \PHI{n}{y},
   }
for values of~$z \notin [0,\,1]$.
This includes~$z$ complex, corresponding to the general case where
one or more of the lines of the meson-quark vertex is not on-shell
in its physical region.
$\GAMMA{n}{z}$ is analytic in the complex plane, with a cut on the
real axis from 0 to~1.
When $x \in [0,\,1]$, $\GAMMA{n}{x}$ is defined by the
principal value prescription, and
   \eqn\vertexeq{
   \GAMMA{n}{x}
   = \left(-\mu_n^2 + { M^2-1  \over x}+{m^2-1\over 1-x} \right)
   \PHI{n}{x}
   ,}
in accordance with~$\BS$.
Since $\PHI{n}{x}$ is finite, $\GAMMA{n}{x}$ has zeros where the first
factor on the right vanishes, and these are the values $x_{\pm}$
where the quarks would be on-shell.  These zeros of the vertex function
cancel quark poles in the propagators of loop amplitudes to ensure
that no quark singularities appear in gauge-invariant Green functions.

All loop integrations are simplified by
the fact that $\PHI{n}{x}$ is a function of $x=p_-/q_-$ only and
is independent of $p_+$.
When wave functions and propagators appear in a loop integrals,
only the latter depend on $p_+$, so the
$\int\d p_+$ is over rational functions and can be computed
explicitly by contour integration, leaving a single integral over
one real variable.

\newsec{Calculating the matrix elements}

In this section we calculate the form factors $f_\pm$ as defined in
Eq.~\ffsdefd.
The current matrix element is given by
\eqn\VERTEXVEV{
\eqalign{
   \vev{\pi |V_\mu|B}
   = {2\over \pi} & \int \d^2k \, \Phi_0(k,p)\Psi(k-q,p')\hat\Phi_\mu(k,q) \cr
   & \qquad\qquad \times S(k)S(k-q)S(k-p) ~, \cr
}}
where $\Phi_0$ is the $B$-meson wavefunction,
$\Psi$ denotes the pion wavefunction
and $\hat\Phi_\mu$ is the full current-quark-quark
vertex, including all resummations of gluon exchange\einhorn.
This vertex function $\hat\Phi_\mu$ has
an intuitive meaning in the physical channel.
It can be expressed as a sum over a complete set of resonances which
couple directly to the current.
This is evident from the formula for the Green functions.
With~$x=k_-/q_-$,
\eqn\fullvertex{
   \hat\Phi_\mu(x,q) = \gamma_\mu - \gamma_-\int{\d y \,\d y'\over
   (y-x)^2}
   G(y,y',q^2) \left[ g_{\mu+}-{M m\over2y'(1-y')q^2_-}g_{\mu-
   }\right]
}
and
\eqn\GREEN{
   G(y,y';q^2) = \sum_n {\phi_n(y) \phi_n(y') \over q^2 - \mu_n^2+i\epsilon}
.}
The identification of the current with the interpolating fields
for the mesons will shortly because obvious.

The asymmetry between plus and minus components,
evident in Eq.~\fullvertex, is a consequence of light-cone gauge.
This has made it too difficult in previous studies to compute the
plus component of the current directly.
We have overcome this obstacle and show below how both
plus and minus components can be used with equal ease ---
a necessity for the study of flavor-changing interactions.

For the `good' component of the current we have
\eqn\goodvertex{
   \eqalign{
   \hat\Phi_-(x,q)
   &=\gamma_-\left( 1-\sum_n {\gn \Phi_n(x)\over q^2- \mu_n^2}
   \right) \cr
   &= \gamma_-\left[ \left(q^2 - {M^2-1\over x} - {m^2-1\over1-x}
   \right) \sum_n
   {\gn \phi_n(x)\over q^2- \mu_n^2} \right] \cr
   &= \gamma_- {1\over q^2} \left[ \left(q^2 - {M^2-1\over x} - {m^2-1\over1-x}
   \right)
   \left( 1 + \sum_n {\gn \mu_n^2 \phi_n(x)\over q^2- \mu_n^2}\right)\right]\cr
}
}
where the last equality holds for $0 \le x \le1$.
The decay constant
$\gn = \int_0^1\d x \phi_n(x)$, should properly be
multiplied by a factor~$\sqrt{\Nc/\pi}$ which we absorb into the
overall matrix element normalization where the $\Nc$-dependence cancels.

A similar expression of the `bad' plus component of the current has
not been presented before, presumably because of the apparent complications
from the associated extra factors in~\fullvertex.
Yet things are not as bad as they look.
The bound-state equation and the parity relation give
two important results\einhorn,
\eqn\parityrelation{
   \int_0^1\d y'\, {M m\over y'(1-y')} \phi_n(y')
   = (-1)^{n}\mu_n^2 \int_0^1\d y'\, \phi_n(y')
   \equiv (-1)^{n}\mu_n^2\gn ~,
}
and
\eqn\GEQ{
   \int_0^1 {\d y \over (y-x)^2} G(y,y';q^2)
   = \delta(y'-x) - \left(q^2 - {M^2-1\over x}-{m^2-1\over1-x}\right)
   G(x,y';q^2) ~.
}

Applying these to Eq.~\fullvertex\ gives
\eqna\eqone
$$
\eqalignno{
   \hat\Phi_+ &=\gamma_++\gamma_-\sum_n(- 1)^n
   {\gn\mu_n^2\over2 q_-^2(q^2-\mu_n)^2 }\Phi_n(x) ~,
   & \eqone a \cr
   \noalign{\hbox{and for $0\le x\le1$,}}
   &= \gamma_+ + \gamma_- {1\over2q_-^2} {M m\over x(1-x)}
   \cr &\qquad
   -\gamma_-{1\over2q_-^2}\left(q^2-{M^2-1\over x}-{m^2-1\over1- x}\right)
   \sum_n(-1)^n{\gn\mu_n^2\phi_n(x) \over q^2-\mu_n^2 }
   &\eqone b ~. \cr
}
$$
Examining the final vertex expressions in
Eqs.~\goodvertex\ and \eqone b,
each contains a weighted sum of pole terms
plus some inhomogeneous, non-pole terms.
The pole terms are of the same form and are readily
interpreted as the contribution from the entire tower of resonant
intermediate states.
The inhomogeneous terms, however, have no such meaning;
happily, they cancel inside matrix elements.
In $\hat \Phi_+$, the $ \gamma_+$ sandwiched
between propagators precisely cancels the second term.
In $\hat\Phi_-$, the term unity, which appears in addition
to the sum would generally
be expected to give rise to a smooth background, quark-model type
contribution, as it arises from a coupling of the bare current
to the valence quark.  However, it drops out of the form
factors since it gives a contribution in $f_\pm$
proportional to
\eqn\CANCEL{
  {2\over q^2 \omega - \mu_B^2\omega}
  \left[  {1\over 1-\omega} \int_0^\omega \d v\phi_0({v\over\omega})
  \Psi( {v-\omega \over 1-\omega} )
   - {1\over\omega} \int_\omega^1 \d v \Phi_0({v\over\omega})
  \psi( {v-\omega \over 1-\omega} ) \right]
.}
This is zero for all $q^2$ since
the two integrals cancel each other.  To see this,
let $v = t \omega$ in the first and let $v = \omega + u(1-\omega)$
in the second, and rewrite each of the $\Phi(x)=\int \d y\, \phi(y)/(y-x)^2$
as a Hilbert transform.
Then the first integral and minus the second are each equal to
\eqn\EQL{
   {\omega \over 1-\omega} \int_0^1 \d t \int_0^1 \d u
   { \phi_0(t) \psi(u)   \over (u - (t-1){\omega \over 1-\omega })^2 }
.}
Therefore the matrix element of the (+)~component of the current
involves a sum over the same wavefunctions as
for the ($-$)~component, with alternating signs.

It is now easy to combine the currents in the invariant
combinations $q\cdot  V$ and $\varepsilon^{\mu\nu}q_\mu V_\nu$,
which couple to states of even and odd parity, or
odd and even~$n$, respectively:
\eqn\invmatels{
   \eqalign{
   \vev{\PI | q^\mu V_\mu | B} &= \vev{\PI(p') | q_+ V_-+q_-V_+ | B(p)}
   = (\mu_B^2-\mu_\PI^2) f_+(q^2) + q^2 f_-(q^2)\cr
   \vev{\PI | q^\mu A_\mu | B} &= \vev{\PI(p') | q_+ V_--q_-V_+ | B(p)}
   = \left( 2\varepsilon^{\mu\nu} p_\mu p'_\nu \right) \, f_+(q^2)
   .\cr}
}
Using Eqs.~\goodvertex, \eqone b,
\eqn\COMB{
   q_+ \hat\Phi_- \pm q_- \hat\Phi_+
   = \gamma_-{1\over 2q_-}\left(q^2-{M^2-1\over x}-{m^2-1\over1- x}\right)
   \sum_{{\scriptstyle n\ {\rm odd}} \atop {\scriptstyle n\ {\rm even}}}
   {\gn\mu_n^2\phi_n(x) \over q^2-\mu_n^2 } ,
}
and applying to Eqs.~\invmatels\ and \VERTEXVEV\ we have
\eqn\fplusgiven{
   f_+(q^2)
   =\sum_{{\rm even}~n}{\gn \gpiBB(q^2) \over q^2 - \mu_n^2}
   \equiv \sum_{{\rm even}\ n}{\A_n(q^2)\over1- q^2/\mu_n^2}
.}
Here
\eqn\Aresidue{
   \eqalign{
   \gpiBB (q^2)={- 2 \mu_n^2 \over  q^2\o - \mu_B^2/\o }
   &\left[{1\over1-\o}\int_0^\o
   {\rm d}v
   \phi_n(v) \phi_0\voo \Psi\vmo   \right.  \cr
   &- \left. {1\over\o}\int_\o^1{\rm d}v \phi_n(v) \Phi_0\voo
   \psi\vmo\right]
   \cr }
   }
and
\eqn\ADEF{
   \A_n(q^2) \equiv - { \gn\gpiBB(q^2) \over \mu_n^2 } ~.
}
The $\gpiBB(q^2)$ are the invariant three-point couplings.
For $q^2$ above threshold
$\omega=p_-/q_-$ is real, between 0 and 1,
\eqn\ogiven{
\o(q^2)={1\over2}\left( 1 + {\mu_B^2-\mu_\PI^2\over q^2} -
\sqrt{ 1 - 2\left({\mu_B^2+\mu_\PI^2\over q^2} \right)
+\left({\mu_B^2-\mu_\PI^2\over q^2}\right)^2}\right)~.
}

For even parity,
\eqn\ODDP{
   \vev{\pi| q^\mu V_\mu|B}
   = \sum_{n\ {\rm odd }} { \gn \gpiBB(q^2) \over q^2 - \mu_n^2 } ~,
}
so the second form factor is given by
\eqn\fminusgiven{
   f_-(q^2)
   ={1\over q^2}\left[
   \sum_{n\ {\rm odd }}{\B_n(q^2)\over1-q^2/\mu_n^2}
   -\sum_{n\ {\rm even}}{\A_n(q^2)\over1-q^2/\mu_n^2}
   \left({\mu_B^2-\mu_\PI^2 }\right)
   \right]
}
where
\eqn\Bresidue{
\eqalign{
   \gpiBB(q^2)={-2\mu_n^2 }
   &\left[{1\over1-\o}\int_0^\o
   {\rm d}v
   \phi_n(v) \phi_0\voo \Psi\vmo   \right.  \cr
   &- \left. {1\over\o}\int_\o^1{\rm d}v \phi_n(v) \Phi_0\voo
   \psi\vmo\right],\cr
}
}
and ${\cal B}_n \equiv - \gn\gpiBB / \mu_n^2$ for $n$~odd.
The $\gpiBB$ have different kinematic prefactors for
even and odd parity as expected
from the Lorentz-invariant couplings of Eqs.~\Leven\
and \Lodd.

Using Cauchy's theorem, we may write
\eqn\CAUCHY{
   f_+(q^2)=\sum_{n\ {\rm even}}{\A_n(q^2)\over1- q^2/\mu_n^2}
   =\sum_{n\ {\rm even}}{\A_n(\mu_n^2)\over1- q^2/\mu_n^2}
.}
The needed convergence of the sum is provided by the
direct analysis of the wavefunctions in Ref.~\einhorn,
which can be carried immediately to the present case to give
that $f\sim 1/|q^2|^{1+\beta_m}$, and $\beta_m > 0$ is defined
in~\Betadef.
The right hand side is far easier to compute, as the residues of
the poles are numbers rather than functions.
What is their meaning?

The residues of the poles are precisely related to the three-point
coupling functions one needs in order to construct an effective
meson Lagrangian.
As in Eq.~\Ldef,
we introduce a field for each meson state and proceed to write down
interaction terms which will reproduce the $S$-matrix elements
computed from 't~Hooft's Feynman rules.

The effective substitution of the on-shell values in \CAUCHY\
corresponds to making non-linear field redefinitions
of the meson states in order to trade higher-derivative interactions
(which arise from Taylor expanding the 3-point functions) for
higher-point interactions.  But the higher order terms---four-point,
five-point, etc.---are already suppressed by extra factors of $\Nc$.
We can choose fields, therefore, in such a way that the only interaction
terms are cubic couplings with no derivatives in the
large $\Nc$ limit.
It is interesting to compare these field redefinitions with
those studied by Georgi in deriving an on-shell Lagrangian
for general low-energy effective field theories\georgi.

\vfill\eject

\newsec{ Numerical Results}

The method used for numerical solution of the three-point couplings
was introduced in Ref.~\JM, to which we refer the reader for
details.
The wavefunctions were expanded in an appropriate Fourier series
and the wave function overlaps computed numerically.
The formulas of the preceding section do not hold for the
lowest residue of each tower so these may be fit from sum rules
along the lines of Ref.~\JM:
For {\it any chosen $q^2 > (\mu_B + \mu_\pi)^2 $ \/},
every term in Eq.~\CAUCHY, can be computed directly except
for~${\cal A}_0(\mu_0^2)$.
Eq.~\CAUCHY\ can be solved for ~${\cal A}_0(\mu_0^2)$ and selecting
different values of~$q^2$
provides an arbitrary number of checks on both the accuracy
and the sum rule.
We find in every case the variation to be in the
fourth significant  figure at most for $q^2$
is chosen to be of the order of the ground  state mass
times a factor of order unity.

We have studied two values of the light-quark mass,
$m=0.56$ and $m=0.1$ ---
corresponding to $\mu_\pi^2=3.09$ and $\mu_\pi^2=0.72$ respectively ---
and numerous values of the heavy-quark mass~$M$.
The coupling constants $\gpiBB$
depend sensitively on the light-quark mass but
weakly on the heavy quark mass.

Figures~1 and~2 show the approach to the chiral limit.
${\cal A}_n(\mu_n^2)$ is plotted against the light quark mass~$m$
for the lowest eight resonances
and it is evident that ${\cal A}_n$ goes to zero for all~$n\ne 0$.
The ground state coupling, ${\cal A}_0$, approaches a constant.
Recall that ${\cal A}_n = - \gpiBB \gn / \mu_n^2$.
Since $\gn$ and $\mu_n^2$ have negligible~$m$ dependence,
the shape is identical to that for $\gpiBB$
as well.

These figures also show the lowest residue clearly dominating the
higher resonances.

\INSERTFIG{
\epsfbox{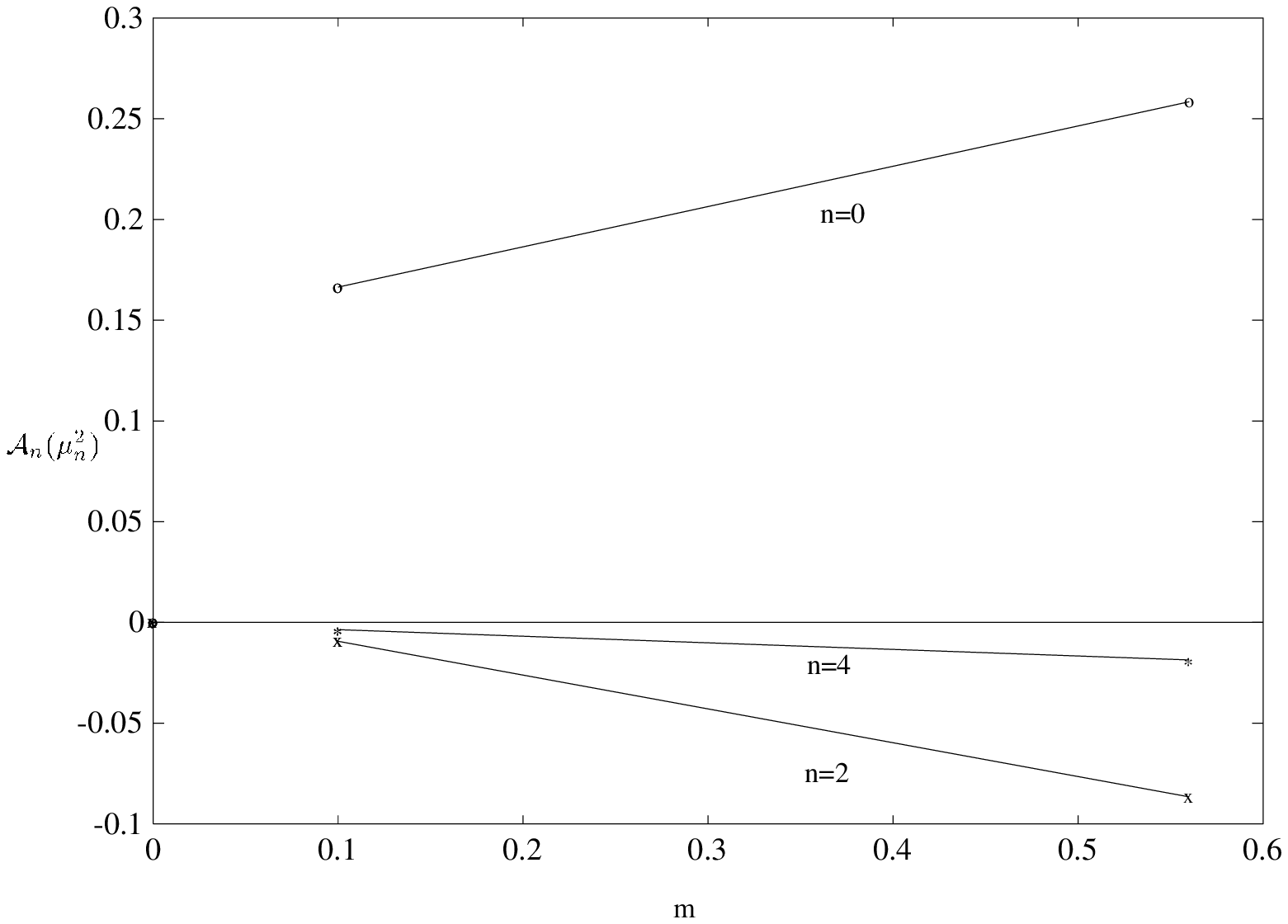}
\CAPTION{Figure 1.}{
The residue $\A_n(\mu_n^2)$ vs. $m$, the light quark mass.
The value was computed numerically for $m=0.1, 0.56$
and a line connecting the pairs of points drawn to guide the eye.
For $n \ne 0$, the $\A_n \to 0$ in the chiral limit, $m\to 0$.
}
}

\INSERTFIG{
\epsfbox{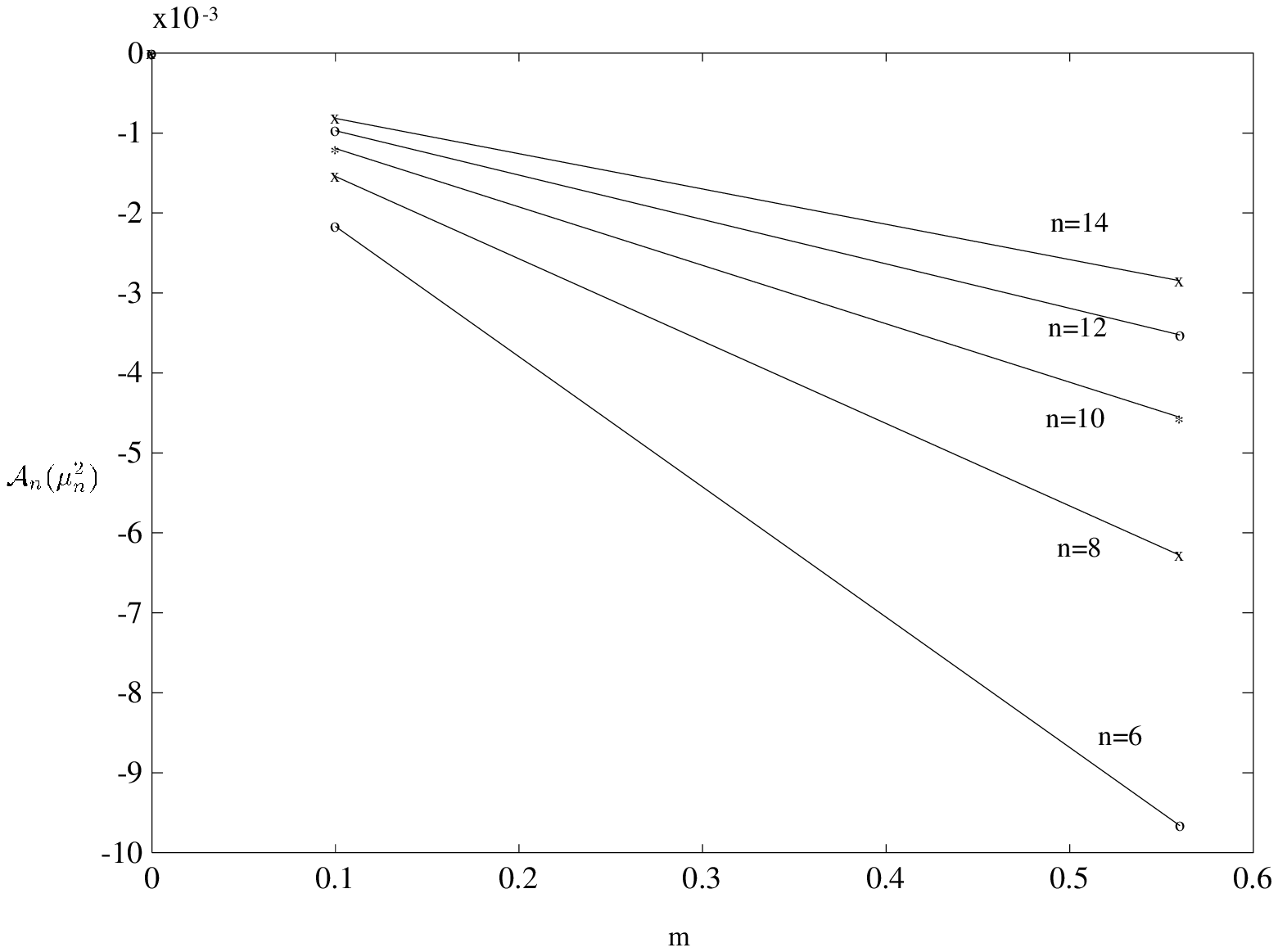}
\CAPTION{Figure 2.}{
The residue $\A_n(\mu_n^2)$ vs. $m$, the light quark mass,
for more resonant states, $n=$6--18.
}
}

Figure~3 shows how the couplings~$\hat \gpiBB \equiv \gpiBB/\mu_n^2$
reach constant values independent
of the heavy quark mass~$M$.

\INSERTFIG{
\epsfbox{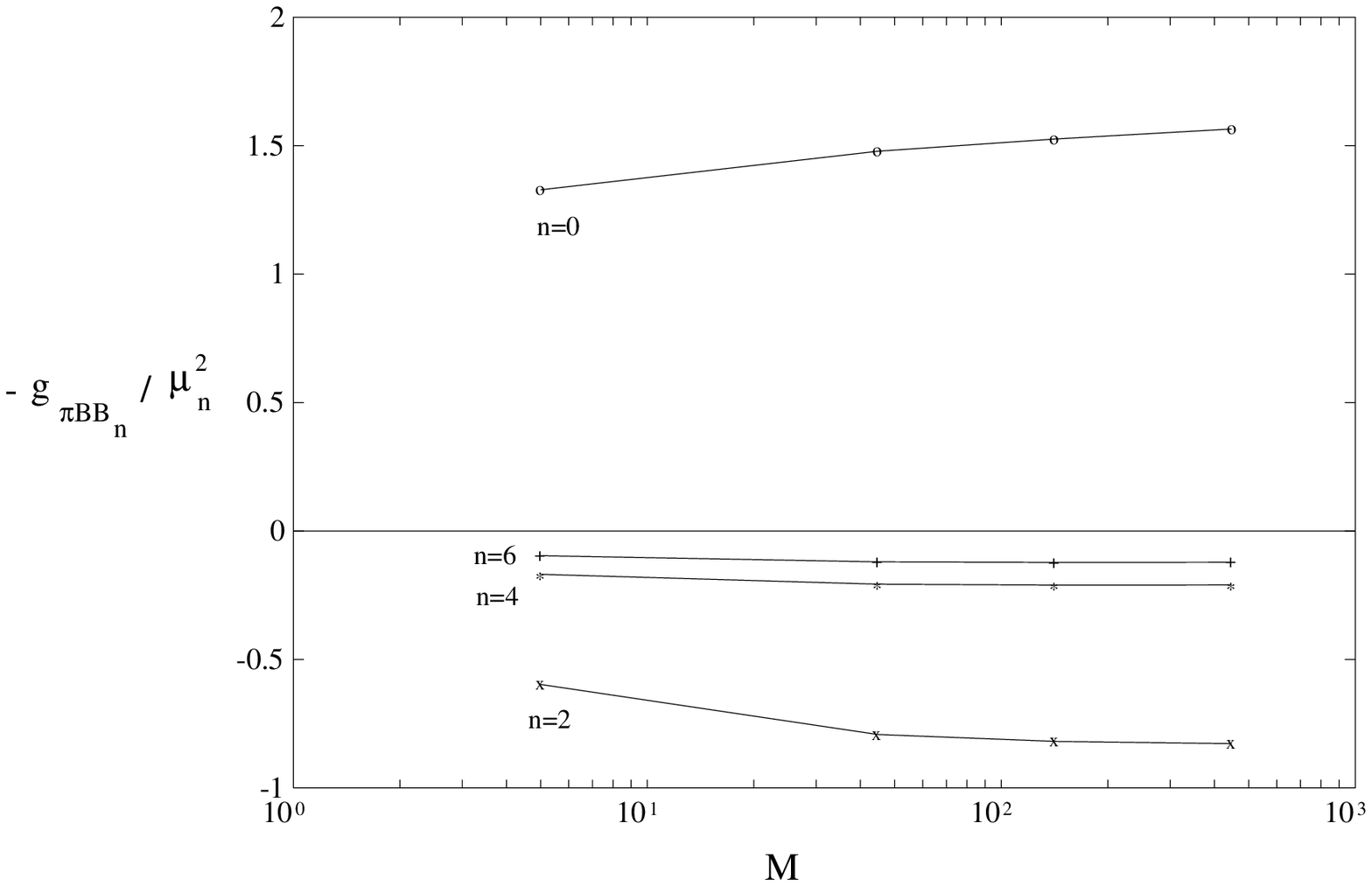}
\CAPTION{Figure 3.}{
Approach to the heavy quark limit:
$-\gpiBB/\mu_n^2$ vs. $M$ for $m=0.56$ ($\mu_\pi^2=3.09$)
and $M^2=25, 2000, 20000, 200000$.
Results for $m=0.1$ are similar.
}
}

While the large coupling $g_{\pi B B^*}$
for the $B^*$ is sufficient for it to dominate,
the factors of the meson decay constants, $\gn$,
further amplify the effect on the form factor.
Figure~4 shows the rapid fall of the~$\gn$ with~$n$.

Finally,
Figures~5 and 6 show the couplings vs. $n$, the state number.
This again illustrates that the lowest pole dominates and
that the pole dominance is stronger as the pion mass decreases,
in agreement with the conclusions of section~3.

\INSERTFIG{
\epsfbox{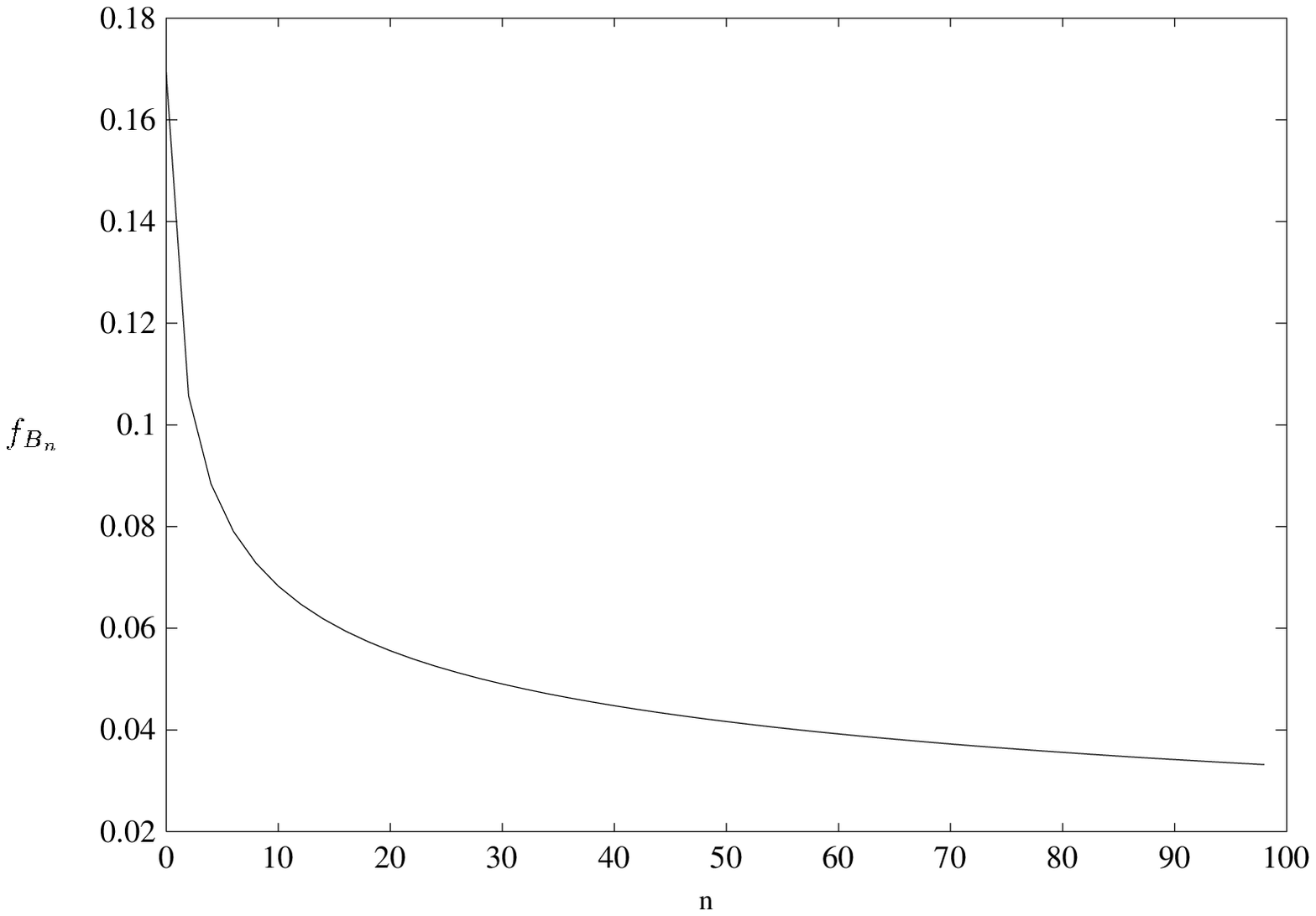}
\CAPTION{Figure 4.}{
Typical $n$-dependence of the decay constants~$\gn$.
(Shown here for $m=0.56$, $M^2=20000$.)
}
}

\INSERTFIG{
\epsfbox{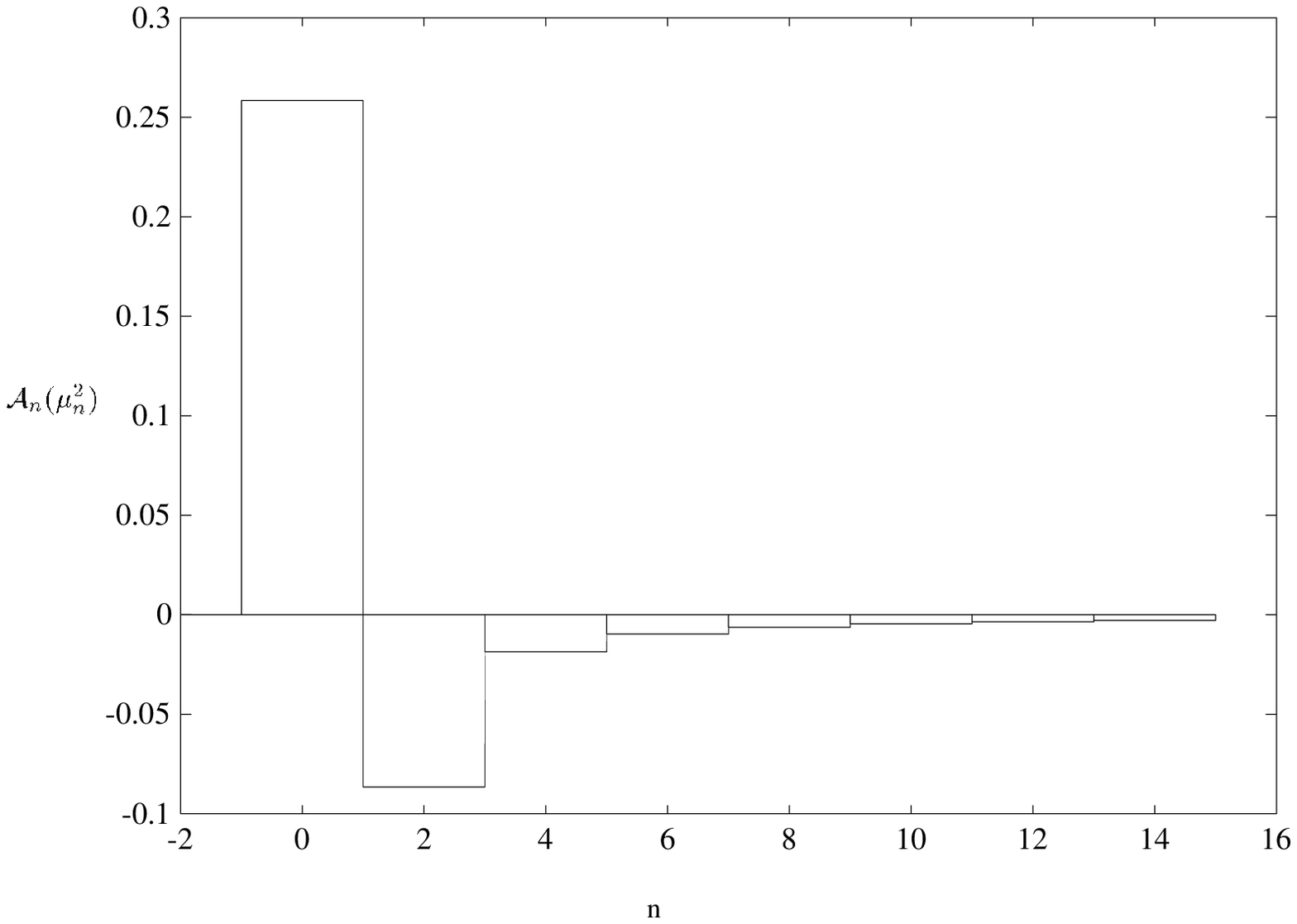}
\CAPTION{Figure 5.}{
Typical $n$-dependence of the pole residues $\A_n(\mu_n^2)$.
(Shown here for $m=0.56$, $M^2=20000$.)
}
}

\INSERTFIG{
\epsfbox{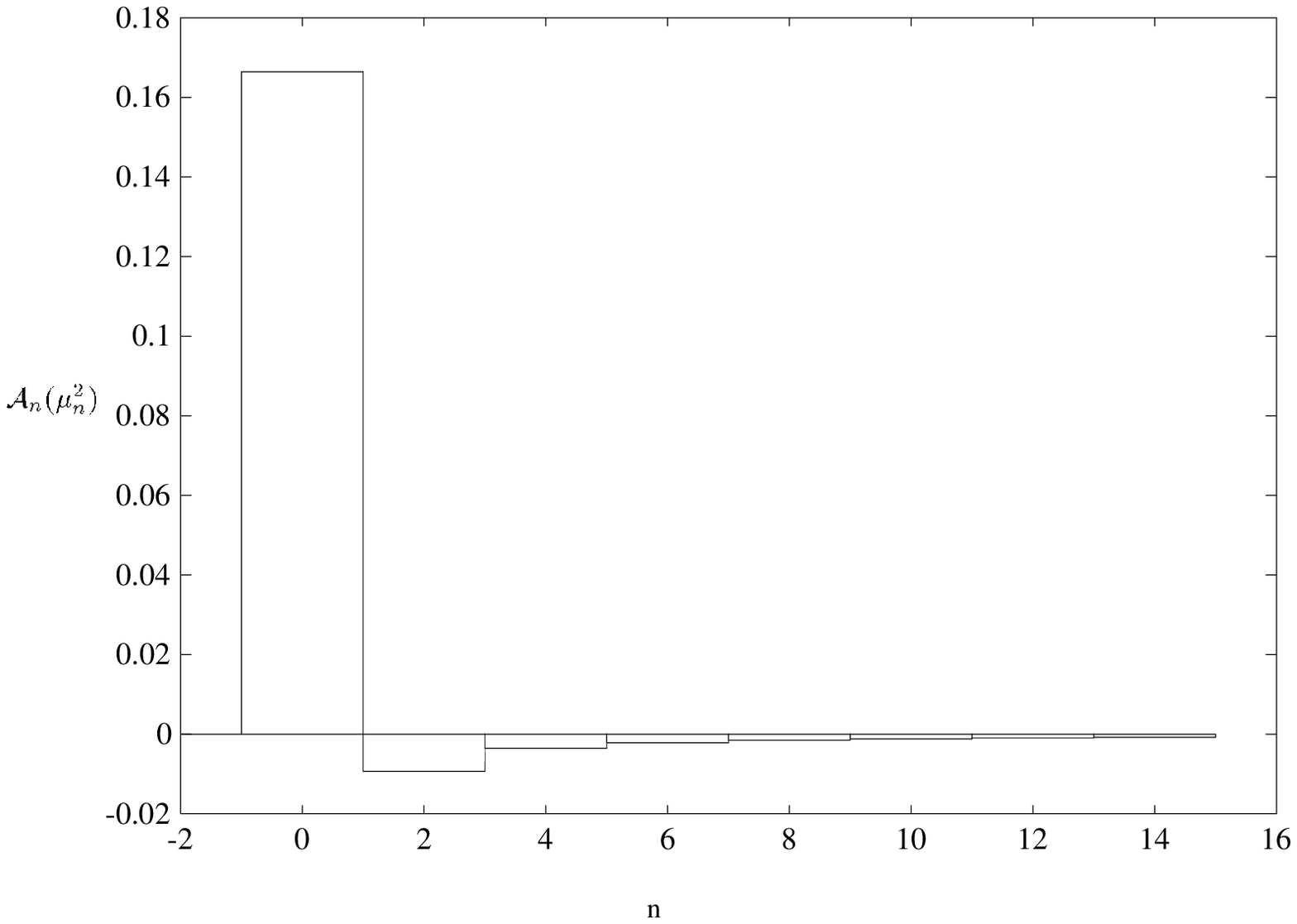}
\CAPTION{Figure 6.}{
Typical $n$-dependence of the pole residues $\A_n(\mu_n^2)$.
(Shown here for $m=0.1$, $M^2=20000$.)
}
}

The $B^*$ pole dominates over a large kinematic region all the way
down to $q^2=0$, not simply the small neighborhood $q^2\approx q^2_{\rm max}$
suggested in Refs.~\IW, \BD.

\newsec{Conclusions}

The decay $\bar B \to \pi e \bar\nu_e$ is dominated by the~$B^*$
vector meson pole over a large kinematic range.
In four dimensions, the result follows from
the combined chiral, heavy-quark, large-$\Nc$ limit.
This provides a justification for the pole-dominated shape
assumed at all~$q^2$ by many model calculations.
In contrast with previous
arguments we find no evidence of a quark-model regime or the need for
two-component models of form factors. The reason for the difference
may be that these arguments assumed a smooth chiral behavior in which
the pion mass could be neglected, whereas we have seen a sensitive
dependence on the pion mass which is in any event of order of the
splitting between states and shouldn't be expected to be negligible
compared to other relevant quantities.

The next step is to use this pole-dominance result for
extracting predictions for the KM mixing angle~$V_{u b}$.
This requires four-dimensional estimates of the corrections
to this limit.  Work in this direction is in progress.
If the corrections turn out to be small it may be possible
to apply these results to semileptonic decays of $D$~mesons,
for $D\to \pi$ or even $D\to K$.

In two dimensions, we have shown that for any fixed heavy quark mass
the form factors for $\bar B \to \pi e \bar\nu_e$ are given by a
single pole at the~$B$. The result was derived in the combined large
$\Nc$ and chiral limit. To study the approach to the chiral limit we
wrote the form factors as resonant sums, gave explicit formulae for
the residues in terms of overlaps of 't~Hooft wave-functions and
computed these overlaps numerically. This demonstrated  the expected
behavior with decreasing pion mass and the expected scaling at large
heavy quark mass.

The nature of deviations from the large $\Nc$ limit should be more
easily addressed in two than in four dimensions.
Also, there is no reason to expect that the
form factors for $\bar B \to \pi^* e \bar\nu_e$, where $\pi^*$ is an
excited pion resonance, should be pole dominated.
Work on these issues is in progress.

{\bf Acknowledgements:}
It is a pleasure to thank
William Bardeen, Sidney Coleman, Roger Dashen, David Gross,
Chung-I Tan and Mark Wise
for helpful conversations.

\listrefs

\vfill\eject

{\bf FIGURE CAPTIONS}

\medskip \item {Fig.\ 1. } %
{The residue $\A _n(\mu _n^2)$ vs. $m$, the light quark mass. The value was
computed numerically for $m=0.1, 0.56$ and a line connecting the pairs of
points drawn to guide the eye. For $n \ne 0$, the $\A _n \to 0$ in the chiral
limit, $m\to 0$.  \medskip }
\medskip \item {Fig.\ 2. } %
{The residue $\A _n(\mu _n^2)$ vs. $m$, the light quark mass, for more
resonant states, $n=$6--18.  \medskip }
\medskip \item {Fig.\ 3. } %
{Approach to the heavy quark limit: $-\gpiBB /\mu _n^2$ vs. $M$ for $m=0.56$
($\mu _\pi ^2=3.09$) and $M^2=25, 2000, 20000, 200000$.
Results for $m=0.1$ are similar.  \medskip }
\medskip \item {Fig.\ 4. } %
{Typical $n$-dependence of the decay constants~$\gn $. (Shown here for
$m=0.56$, $M^2=20000$.)  \medskip }
\medskip \item {Fig.\ 5. } %
{Typical $n$-dependence of the pole residues $\A _n(\mu _n^2)$.
(Shown here for
$m=0.56$, $M^2=20000$.)  \medskip }
\medskip \item {Fig.\ 6. } %
{Typical $n$-dependence of the pole residues $\A _n(\mu _n^2)$.
(Shown here for
$m=0.1$, $M^2=20000$.)  \medskip }

\bye